\documentclass[12pt]{article}
\usepackage{booktabs}
\usepackage[authoryear]{natbib}
\usepackage[margin=2cm,a4paper]{geometry} 
\usepackage{amssymb, amsthm, amsmath}
\usepackage[pdftex]{graphicx}
\usepackage{subfig}
\usepackage{natbib}

\begin{document}

\newcommand{\fra}[2]{\textstyle{\frac{#1}{#2}}}

\newcommand{\pf}{\noindent{\em Proof: }}
\newcommand{\epf}{\hfill\hbox{\rule{3pt}{6pt}}\\}
\newtheorem{theorem}{Theorem}
\newtheorem{corollary}{Corollary}

\newcommand{\beqn}{\begin{eqnarray}\begin{aligned}}
\newcommand{\eqn}{\end{aligned}\end{eqnarray}}


\title{Low-parameter phylogenetic estimation under the general Markov model}

\author{Barbara R. Holland, Peter D. Jarvis, and Jeremy G. Sumner}
\maketitle

\begin{abstract}\noindent
In their 2008 and 2009 papers, Sumner and colleagues introduced the ``squangles'' -- a small set of Markov invariants for phylogenetic quartets. 
The squangles are consistent with the general Markov model (GM) and can be used to infer quartets without the need to explicitly estimate all parameters.
As GM is inhomogeneous and hence non-stationary, the squangles are expected to perform well compared to standard approaches when there are changes in base-composition amongst species.
However, GM includes the IID assumption, so the squangles should be confounded by data generated with  invariant sites or with rate-variation across sites. 
Here we implement the squangles in a least-squares setting that returns quartets weighted by either confidence or internal edge lengths; and use these as input into a variety of quartet-based supertree methods.
For the first time, we quantitatively investigate the robustness of the squangles to the breaking of IID assumptions on both simulated and real data sets; and we suggest a modification that improves the performance of the squangles in the presence of invariant sites.
Our conclusion is that the squangles provide a novel tool for phylogenetic estimation that is complementary to methods that explicitly account for rate-variation across sites, but rely on homogeneous -- and hence stationary -- models.
\end{abstract}

\footnotesize{\noindent
School of Mathematics and Physics, University of Tasmania, Australia\\
\textit{keywords:} Markov invariants, phylogenetic invariants, base-composition, rate-variation, quartets, supertrees\\
\textit{email:} barbara.holland@utas.edu.au}
\normalsize

\section*{Introduction}

There is a consensus opinion that the most robust and easily interpretable phylogenetic methods are based on stochastic models of how sequences evolve. 
Many simulation studies comparing phylogenetic methods have supported the use of model-based likelihood methods over parsimony \citep{Hi1995, Hu1995, GuGa2003, GaMi2005}.
However, the question of whether current models of sequence evolution are adequate is still an open one.

The continuous-time Markov models of nucleotide evolution in common use -- namely the general-time-reversible model (GTR) and its submodels -- are stationary, reversible and homogeneous \citep{Je2008}.
Stationarity implies that the expected base-composition should be constant across the tree.
However, in real data significant differences in base-composition have been observed \citep{Lo1992,FoHi1999,PhPe2003,Gr2007}, and it has been shown that species with similar base-composition tend to group together regardless of their evolutionary history \citep{Je2004,PhDePe2004,DaPe2008} -- although the base-composition bias may have to be quite pronounced before the accuracy of maximum-likelihood (\emph{ML}) significantly diminishes \citep{GaGo1995}.
If the true evolutionary process was heterogeneous across the tree, \citet{sumner2011,sumner2012a} have shown that GTR has an undesirable property that makes a consistent interpretation difficult.

To accurately capture biological processes, more parameter-rich models than the GTR family may sometimes be required.
For example, \citet{GaGo1995,GaGo1998} explore models that incorporate differences in base-composition.
Other researchers have discussed using mixtures of rate matrices from the GTR class, either as mixtures across the tree \citep{PaMe2004}, or as a temporal hidden Markov model where different rate matrices may apply in different parts of tree \citep{Wh2008}. 
\citet{BaHa1987b,BaHa1987a}  suggested the general Markov model (GM) for use in phylogenetics. 
This model allows distinct Markov matrices to be assigned to each edge in the tree, with only the restriction that the entries of the matrix be interpretable as substitution probabilities.

Implementing GM for phylogenetic inference is problematic because it is very parameter rich.
Even for just three taxa, there are 39 free parameters; and each additional taxon presents another 24 parameters to be estimated.
For a homogeneous implementation of GTR the corresponding count is nine free parameters with each additional taxon requiring the addition of only two more. 
Models that are very parameter rich lose statistical power: ``... extensive addition of parameters comes at a price -- the predictive power of the theory (the information that the data can reveal about the underlying tree) tends to be drowned out in a sea of parameter estimation'' \citep{St2005}.
This is the well-known statistical issue of bias-variance tradeoff; where, for parameter-rich models, parameter estimates may be relatively unbiased but will have relatively high variance \citep{burnham2002}.

The worst case for the bias-variance tradeoff occurs when a phylogenetic model is not ``identifiable''.
Identifiability ensures that the parameter values that are input to the model can be computed exactly from the resulting probability distribution of site patterns.
If a model is not identifiable, then there are parameters in the model whose estimation variance is formally infinite.
\citet{Ch1996} proved that GM is identifiable and \citet{AlRh2008} extended the result to include invariant sites (although both results are contingent on potential ``label swapping'' of character states at the internal vertices of the phylogenetic tree).

There has been interest in implementing both GM \citep{JaJeRo2005} and GM+I \citep{JaRoJe2007} in an \emph{ML} context.
By conducting a simulation study on triples, \citet{Os2008} compared five different methods for phylogenetic estimation under GM;
their results suggested that the methods presented in \citet{Go1996} and \citet{Go1982} -- as implemented in \citet{Kn2007} -- provide a reasonable compromise between computational efficiency and statistical accuracy.
In contrast, \citet{Zo2011} suggest that the ``label swapping'' proviso causes heuristic search methods to frequently get stuck in parameter regions that correspond to sub-optimal likelihood values.

Taking these considerations into account, it is of interest to develop phylogenetic methods that are both consistent with more general models of sequence evolution and avoid the dangers of over-parametrisation.
The first advance in this regard was the logDet distance \citep{Lo1994, La1994, BaHa1987b}. 
A distance between two taxa $x$ and $y$ can be computed as $d_{xy}\!=\! -\log(\det(F_{xy}))$ where $F_{xy}$ is the $4\times 4$ divergence matrix for the sequences $x$ and $y$.
For example, the matrix element $\left(F_{xy}\right)_{AC}$  counts the number of sites where an A was observed in sequence $x$ and a C was observed for sequence $y$.
This distance can be used to consistently estimate tree topology, although as noted in \citet{Lo1994} if a consistent estimate of edge lengths is desired then the formula requires an extra term involving the base-composition at each of the taxa.
These pairwise distances can then be used to build large trees by using distance-based methods such as Neighbour Joining (\emph{NJ}).
Bypassing the need to estimate a complete set of model parameters, \emph{logDet+NJ} consistently estimates tree topology under GM \citep{steel1994} -- thus providing a sensible compromise to the bias-variance tradeoff.


The logDet function is constructed from the simplest example of a ``Markov invariant'' \citep{Su2008}. 
From this point of view, the defining feature for the logDet function is the following property of $\det(P_{xy})$ -- where the divergence matrix $F_{xy}$ has been replaced with the corresponding theoretical probability distribution $P_{xy}$ arising under GM.
Consider an edge $e$ that lies on the path between taxa $x$ and $y$ with associated Markov matrix $M_e$.
If we extend this edge by inserting an additional Markov matrix $M_e'$ so that $M_e\rightarrow M_e'M_e$, then the consequent change in $\det(P_{xy})$ is given by $\det(P_{xy})\rightarrow \det(M'_e)\det(P_{xy})$.
Recall that in a continuous-time formulation, $M'_e=e^{Q'_et'_e}$ for some rate matrix $Q'_e$ and time $t'_e$, so $-\log(\det(M'_e))=\left(\text{sum of rates in }Q_e'\right) t'$; whence the interpretation of the logDet as a distance measure consistent with GM.

The theory of Markov invariants applies exactly this idea to larger subsets of taxa: polynomial functions of phylogenetic divergence arrays that are multiplied by a power of the determinant of the Markov matrix that extends a pendant edge.
The restriction to \emph{pendant} edges is not relevant for the logDet but is pivotal in the general case.
\citet{Su2008} formalise this definition and describe methods to calculate the number of Markov invariants that exist for a given number of taxa and character states.
In particular, for quartet trees under GM on four character states -- i.e. DNA, they show there are four Markov invariants; and they refer to these as the ``squangles''.
(The term squangle derives from \textbf{s}tochastic \textbf{qu}artet t\textbf{angle}; more explanation is given in \citet{Su2006}.)

For a quartet with pendant edges $\ell_x,\ell_y,\ell_z$ and $\ell_w$, the defining feature of the squangles is their behaviour when these edges are extended by inserting the additional Markov matrices $M_{\ell_x}',M_{\ell_y}',M_{\ell_z}'$ and $M_{\ell_w}'$.
If $q(P_{xyzw})$ is the value of a squangle before the extension, then \[q(P_{xyzw})\rightarrow \det(M_{\ell_x}')\det(M_{\ell_y}')\det(M_{\ell_z}')\det(M_{\ell_w}')q(P_{xyzw}).\]   
It is precisely this behaviour that makes the squangles analogous to the ``Det'' part of the logDet function.

It is important to note that Markov invariants are in general different from phylogenetic invariants \citep{cavender1987,EvSp1993,felsenstein1991,La1987,St1993}.
For a given model of sequence evolution, phylogenetic invariants are polynomial functions that evaluate to zero for a subset of phylogenetic trees regardless of particular model parameters.
\citet{SuJa2009} use the symmetries of leaf permutation on quartets to show that, for each possible quartet topology, the squangles can be arranged into a prescribed basis where two of them form phylogenetic invariants.
This amalgamation of the properties of Markov and phylogenetic invariants is a very special circumstance and is not achievable in general.

Following the investigation of the effect of base-composition bias given in \citep{Je2004}, \citet{Su2008} conducted simulations showing that the squangles can be used to infer quartets accurately  under both conditions of extreme base-composition bias and relatively short sequence lengths (performing at least as well as \emph{logDet+NJ}). 
This is consistent with the results presented in \citet{casanellas2007}, and is in contrast to previous analyses of phylogenetic invariants which have generally shown low-power compared to other phylogenetic methods \citep{HuHi1993}

As with the logDet, the squangles are not designed to handle situations where the data arises under a process that breaks the IID assumptions, i.e. invariant sites or rate-variation across sites. 
For the first time, we explore to what extent the squangles are robust to such violations. 
This kind of ``tool abuse'' is common in phylogenetic practice.
As alluded to above, maximum-likelihood methods are often used in situations where the models are clearly not great matches to reality; and simulations have been used to suggest that accuracy of phylogenetic inference under \emph{ML} is reasonably robust to many kinds of model violation.

In this paper we aim to determine if the squangles can become a practical tool for phylogenetic estimation. 
In particular we address the following questions:
\begin{enumerate}
\item How can we measure our confidence in the quartet tree returned by the squangles?
\item How can we best use the squangles in combination with existing software to perform phylogenetic estimation for many taxa?
\item In what settings would the squangles outperform \emph{NJ} with distances computed assuming a stationary model, \emph{NJ} using logDet distances, or \emph{ML} assuming homogeneous models?
\item How robust are the squangles to the presence of invariant sites? 
\item Can the squangle method be modified to make it robust to invariant sites?
\end{enumerate}

\section*{Methods}

\subsection*{The squangles}

A DNA sequence alignment for four species can contain up to $4^4 = 256$ site patterns (we assume no gaps, missing data or ambiguous characters). 
The relative frequencies (or proportions) of site patterns can be summarised in a vector $\bf f$ of length 256.
The squangles are then homogeneous polynomials of degree 5 in these pattern frequencies. 
Using the basis prescribed in \citet{SuJa2009}, there are three squangles that are useful for phylogenetic quartet estimation: here denoted as $q_1(\bf f)$, $q_2(\bf f)$, and $q_3(\bf f)$.
Each of these has 66,744 terms, and together they satisfy the linear relation $q_1 + q_2 + q_3 = 0$ (which is to say that up to linear dependence there are only two of them).

\citet{SuJa2009} give algebraic expressions for the squangles, and derive their expectation values when the pattern frequency vector $\bf{f}$ arises from GM on a given quartet topology.
These expectation values are given in Table~\ref{tab:squangexp}. 
For instance, if the pattern frequency $\bf{f}$ has been generated on the quartet $12|34$ -- meaning that the middle edge splits taxa 1 and 2 from taxa 3 and 4, $q_1$ has expectation value $E[q_1]=0$, $q_2$ has non-positive expectation value $E[q_2]=-u\leq 0$, and $q_3$ has expectation value $E[q_3]=u$ (equal in magnitude but opposite in sign to the value for $q_2$).
The  numerical value of $u$ will depend on the particular Markov matrices that underlie the process that generates the data.
We will discuss this in detail below; but note here that the symmetries inherit in Table~\ref{tab:squangexp} imply that $u=v=w=0$ only in the case of a star-tree.

\begin{table}[ht]
\begin{center}
\begin{tabular}{c|c|rrr}
Hypothesis & Quartet & $E[q_1]$ & $E[q_2]$ & $E[q_3]$ \\
\hline
$\mathcal{H}_1$ & $12|34$ & 0 & $-u$ & $u$ \\
$\mathcal{H}_2$ & $13|24$ & $v$ & 0 & $-v$ \\
$\mathcal{H}_3$ & $14|23$ & $-w$ & $w$ & 0 \\
\end{tabular}  
\end{center} 
\caption{Expected values of each of the squangles on each of the three possible quartet topologies,
with $u$, $v$, $w \geq 0$.} 
\label{tab:squangexp}
\end{table}

The expectation values of the squangle $q_1$ and the linear combination $q_2+q_3$ are zero on $12|34$.
Hence these are phylogenetic invariants for that quartet (with complementary phylogenetic invariants occurring for the other two quartets).
As can be confirmed  from Table~\ref{tab:squangexp}, the relation $q_1+q_2+q_3=0$ implies that $E[q_1+q_2+q_3]=E[q_1]+E[q_2]+E[q_3]=0$ for each quartet topology.

Recall that \citet{steel1994} used the properties of the logDet distance to prove that (unrooted) tree topology is identifiable under GM.
The existence of the squangles and their properties as presented in Table~\ref{tab:squangexp} provides an independent verification of this result. 

\subsection*{Least-squares implementation of the squangles}

Given an arbitrary site pattern frequency vector $\bf f$, the squangles $q_1(\bf f)$, $q_2(\bf f)$, and $q_3(\bf f)$ will evaluate to values that will not exactly match any of the three scenarios given in Table 1 (they will do so only for infinite sequence lengths and when the process that generated $\bf{f}$ was precisely GM on a fixed quartet). 
Following \citet{Su2008}, we consider each quartet as a distinct hypothesis (see Table~\ref{tab:squangexp}) and use a least-squares framework to decide which hypothesis best explains the observed values of the squangles.
We must be careful to take into consideration the linear relation $q_1+q_2+q_3=0$, which implies that, for each quartet, we should consider only two of the squangles as independent.
To achieve this in a way that is consistent when considered across the three quartets, for each case we ignore the squangle with vanishing expectation value.  
For instance, for the quartet $12|34$, we ignore $q_1$ and define the residual sum of squares as $RSS_1 =  (q_2 +u)^2 + (q_3 - u)^2$.

We want to find the least-squares estimate for $u$ subject to the constraint $u\geq 0$. 
To do so, we substitute $x^2 = u$, differentiate the residual with respect to $x$ and set the result equal to zero:
\[\frac{dRSS_1}{dx} = 4x(q_2+x^2) - 4x(q_3 - x^2) = 0.\]
Solving gives two solutions $u = x^2 = 0$ and $u = x^2 = \frac{1}{2}(q_3 - q_2)$.
By checking the sign of the second derivative we determine that when $q_2 > q_3$ the minimum occurs at $u=0$, and when $q_2 < q_3$ the minimum occurs at $u = \frac{1}{2}(q_3 - q_2)$.
If $q_2 = q_3$ then both solutions are the identical and the minimum is at $u=0$. 
Note that these conditions ensure that $u\geq 0$ (as they must since $u=x^2$).

Applying the complementary procedure to define residuals for the other two quartets -- in each case ignoring the squangle with vanishing expectation value -- we obtain the following least-squares estimates for the parameters $u$, $v$ and $w$ under the hypotheses of the three respective quartets:
\beqn
\hat{u} &= \max\{0, \fra{1}{2}(q_3-q_2)\},\\
\hat{v} &= \max\{0, \fra{1}{2}(q_1-q_3)\},\\
\hat{w} &= \max\{0, \fra{1}{2}(q_2-q_1)\}.\nonumber
\eqn

Given the least-squares estimates of $u$, $v$, or $w$ for each hypothesis, Table~\ref{tab:RSSvalues} displays the corresponding RSS values for each of the 6 possible configurations of the squangles as ordered on the real number line.
We deem the hypothesis with the strongest support to be the quartet, or quartets in the case of ties, with the minimum RSS.

If, as presented in Table~\ref{tab:squangexp}, the observed values of $q_1$, $q_2$, and $q_3$ match the order defined by the expectation values for a particular quartet, then Table~\ref{tab:RSSvalues} shows that the RSS for that quartet will always be lower than the RSS for the other two topologies. 
For example, the configuration $q_2 \leq  q_1 \leq  q_3$ in row 1 of Table~\ref{tab:RSSvalues} matches the order given in row 1 of Table~\ref{tab:squangexp}.

\begin{table}[ht]
  \centering
    \begin{tabular}{lllrllrll}
    \toprule
     & \multicolumn{2}{c}{$\mathcal{H}_1=12 |34$} &\phantom{abc}& \multicolumn{2}{c}{$\mathcal{H}_2=13|24$} &\phantom{abc}& \multicolumn{2}{c}{$\mathcal{H}_3=14|23$}  \\
   \midrule
    Ordering\phantom{ab}     & $\hat{u}$     & RSS$_1$   					&& $\hat{v}$      & RSS   && $\hat{w}$      & RSS \\
\cmidrule{2-3}
\cmidrule{5-6}
\cmidrule{8-9}\noalign{\smallskip}
    $q_2\leq q_1\leq q_3$     & $\frac{1}{2}(q_3-q_2)$     & $\frac{1}{2}q_1^2$     && 0     & $q_1^2 + q_3^2$     					&& 0     & $q_1^2 + q_2^2$  \\
    $q_3\leq q_2\leq q_1$     & 0     			& $q_2^2 + q_3^2$     			&& $\frac{1}{2}(q_1-q_3)$     & $\frac{1}{2}q_2^2$      && 0     &  $q_1^2 + q_2^2$ \\
    $q_1\leq q_3\leq q_2$     & 0     			& $q_2^2 + q_3^2$     			&& 0     & $q_1^2 + q_3^2$     						&& $\frac{1}{2}(q_2-q_1)$     & $\frac{1}{2}q_3^2$ \\
    &&&&&&&&\\
    $q_3\leq q_1\leq q_2$     & 0     			& $q_2^2 + q_3^2$     			&& $\frac{1}{2}(q_1-q_3)$     & $\frac{1}{2}q_2^2$      && $\frac{1}{2}(q_2-q_1)$     & $\frac{1}{2}q_3^2$ \\
    $q_1\leq q_2\leq q_3$     & $\frac{1}{2}(q_3-q_2)$     & $\frac{1}{2}q_1^2$     && 0     &  $q_1^2 + q_3^2$     				 	&& $\frac{1}{2}(q_2-q_1)$     &  $\frac{1}{2}q_3^2$  \\
    $q_2\leq q_3\leq q_1$     & $\frac{1}{2}(q_3-q_2)$     & $\frac{1}{2}q_1^2$     && $\frac{1}{2}(q_1-q_3)$     & $\frac{1}{2}q_2^2$     && 0     &  $q_1^2 + q_2^2$ \\
    \bottomrule
    \end{tabular}
      \caption{Least-squares estimates and corresponding residual sums of squares for each of the three possible quartet hypotheses.
  Note that the first three rows give orderings on the squangles that match the ordering of expectation values presented in Table~\ref{tab:squangexp}.}
  \label{tab:RSSvalues}
\end{table}

If the configuration of $q_1$, $q_2$, and $q_3$ doesn't match one of the orders displayed in Table~\ref{tab:squangexp}, i.e. rows 4--6 of Table~\ref{tab:RSSvalues}, then one hypothesis is excluded -- the one where the ordering is ``completely wrong'' -- and, out of the remaining two hypotheses, the minimum RSS occurs for the hypothesis $\mathcal{H}_i$ whose corresponding squangle $q_i$ is closest to zero.

In the case of a single equality between $q_1$, $q_2$, and $q_3$, the configuration matches \emph{two} of the orders displayed in Table~\ref{tab:RSSvalues}: one matches an order displayed in Table~\ref{tab:squangexp} and one does not.
It is easy to check that the same minimum RSS occurs for both possibilities.   

The only circumstance where the squangles are equal occurs when $q_1 = q_2 = q_3 = 0$.
In this case the RSS values of the hypotheses are also equal.

\begin{figure}[ht]
\centering
\includegraphics[width=40em]{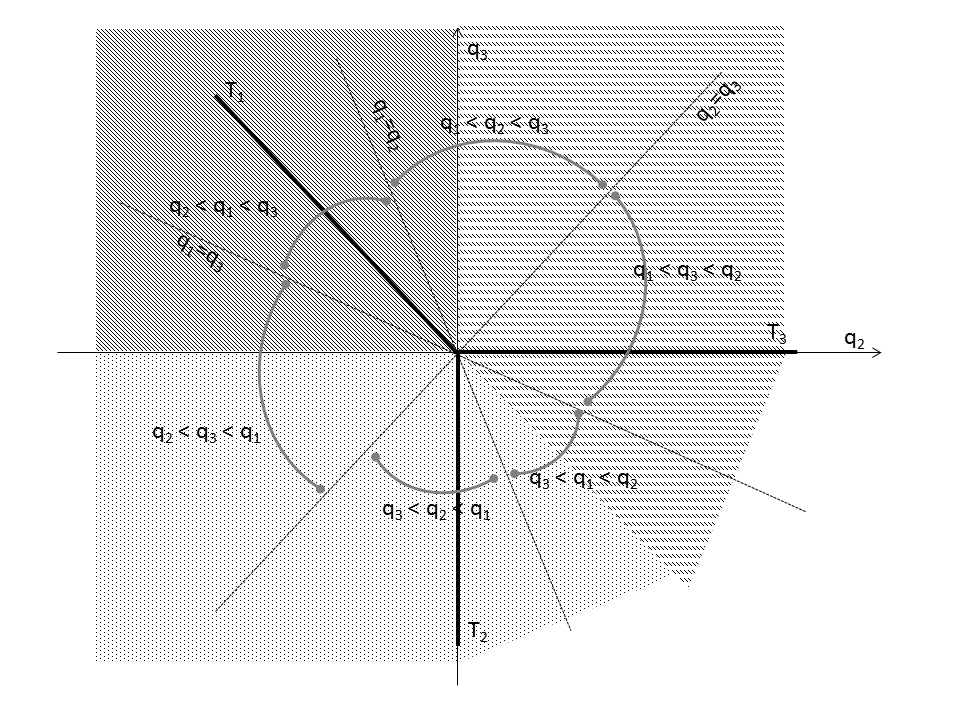}
\caption{Regions in the $q_2$-$q_3$ plane. 
Different regions of the plane correspond to different configurations $q_1$,$q_2$, and $q_3$ (indicated by grey arcs), and to different tree topologies (indicated by shading). 
Bold black lines correspond to squangle values that exactly match one of the three tree topologies.}
\label{fig:q2q3plane}
\end{figure}

Figure~\ref{fig:q2q3plane} summarises the information in Table~\ref{tab:squangexp} as it is interpreted on the $q_2$-$q_3$ plane, showing the  regions where the different configurations apply and the different quartets that are selected.

\subsection*{Accounting for invariant sites}

As GM includes the IID assumption, the squangles are not expected to perform well in cases when the data are generated under a process with significant numbers of invariant sites or when there is rate-variation across sites. 
To account for the possibility of invariant sites, we suggest a simple modification to the calculation of the squangles. 
Firstly, the proportion $\nu$ of invariant sites is estimated using the method given in \citet{St2000}.
Then the observed proportions of constant site patterns -- i.e. the patterns AAAA, CCCC, GGGG and TTTT -- are rescaled by multiplying by $1-\nu$, and the pattern vector ${\bf f}$ is renormalised to sum 1. 
The altered pattern vector ${\bf f}$ is then used as input to the squangle method with no further modification.

\subsection*{Quartet weights}

Several of the methods proposed for constructing trees based on sets of quartets rely on the quartets having weights,
for example, \textit{QuartetSuite} \citep{Wi1999}, \textit{Quartet MaxCut} \citep{SnRa2012} and \textit{QNet} \citep{Gr2008}. 
Although they currently do not require them, presumably supertree methods such as \textit{Matrix Representation with Parsimony} (\textit{MRP}) \citep{Ba1992, Ra1992} or supernetwork methods such as  \textit{Z-Closure} \citep{Hu2004, Hu2006}  or \textit{Q-Imputation} \citep{Ho2007, Ho2008} could also be extended to make use of such weights.
In most cases the weights are required to be proportional to the confidence in a particular quartet topology.
An exception is the program \textit{QNet} \citep{Gr2008}, where the quartet weights are required to be proportional to the
length of the internal edge of the quartet.

For the squangle method implemented using least-squares, we take the RSS values to be inversely proportional to our confidence in the corresponding quartets:
\beqn
w_i &:= \frac{RSS_i^{-1}}{{RSS_1}^{-1}+{RSS_2}^{-1}+{RSS_3}^{-1}},\nonumber
\eqn
with $w_1+w_2+w_3=1$.
This definition can be justified by supposing that, for fixed sequence length and the quartet $12|34$, $q_2$ and $q_3$ are independent and normally distributed around their respective means ($-u$ and $u$) with identical standard deviation $\sigma$. 
Thus the joint probability density function is given by
\beqn
\rho(q_2,q_3;u,\sigma)=\left(\frac{1}{\sqrt{2\pi}\sigma}\right)^2\exp\left(-{\frac{(q_2+u)^2+(q_3-u)^2}{2\sigma^2}}\right).\nonumber
\eqn
Under this assumption, it is clear that the maximum-likelihood estimates are given exactly as $\widehat{u}=\frac{1}{2}(q_3-q_2)$ and $\widehat{\sigma}^2=\fra{RSS_1}{2}=\fra{1}{2}q_1^2=\fra{1}{2}(q_2+q_3)^2$.
As is well known, for these assumptions $\widehat{u}$ is exactly the same as the best estimate arising from the least-squares procedure.
Given observed values $q_2$ and $q_3$, the value of the probability density function at the maximum-likelihood estimates is proportional to the residual sum of squares:
\beqn
\rho(q_2,q_3;\widehat{u},\widehat{\sigma})=\left(\frac{1}{\sqrt{2\pi}\sqrt{\frac{RSS_1}{2}}}\right)^2\exp\left(-{\frac{(q_2+\frac{1}{2}(q_3-q_2))^2+(q_3-\frac{1}{2}(q_3-q_2))^2}{2\frac{RSS_1}{2}}}\right)
=\frac{1}{e\pi}RSS_1^{-1}.\nonumber
\eqn 
From this we conclude that our definition of the weights $w_i$ is consistent with the assumption that the squangles are independent and normally distributed around their respective means with identical variance. 
We present a simulation study below that provides evidence that this assumption is reasonable.
While this gives a plausible theoretical justification of the least-squares approach and the interpretation of the weights $w_i$, we emphasise that our primary aim in this paper is to use simulation and examples from real data to show that the squangles provide a robust phylogenetic tool.

It is apparent from the symmetries in Table~\ref{tab:squangexp}  that $\hat{u}$ (or $\hat{v}$ or $\hat{w}$) must be correlated to the internal edge length of the quartet.
However, as presented in \citet{SuJa2009}, the precise relationship under the full assumptions of GM is too complicated to be useful. 
If we simplify the assumptions of GM on a quartet so on the internal edge we restrict to the Jukes-Cantor (JC) model \citep{jukes1969}, then it is possible to derive an explicit expression for the relationship between the internal edge length and the expectation values $u$, $v$ and $w$.

To achieve this, fix the quartet $12|34$ and suppose that the JC substitution matrix on the internal edge has off-diagonal entries equal to $a/3$.
As is shown in the appendix, the expectation value of the squangle $q_3$ can then be written as a quintic function of $a$:
 \begin{equation}
 E[q_3(a)] = u = \frac{1}{2^73^2}a(3^4-225a+276a^2-154a^3+32a^4)\prod_{i=1}^{4}\det(M_i),
 \label{eq:intedgeval}
 \end{equation}
 where $M_1$ to $M_4$ are GM substitution matrices on the pendant edges.

Using the approach that forms the foundation of the logDet distance, the determinants of the substitution matrices that appear in (\ref{eq:intedgeval}) can be estimated, as follows.
Let $F_{xy}$ be the $4\times 4$ divergence matrix for taxa $x$ and $y$, and let ${P}_{xy}$ be the corresponding theoretical probability distribution under the assumption of GM on pendant edges and JC on the internal edge. 
Assuming the quartet $12|34$, we have $\det({P}_{12}) = \det(M_1)\det(M_2)(\frac{1}{4})^4$ and $\det(P_{34}) =\det(M_3)\det(M_4)(\frac{1}{4})^4$.
Rearranging and substituting this into equation (\ref{eq:intedgeval}) gives
\beqn
\gamma(a)-E\left[\frac{q_3}{\det(P_{12})\det(P_{34})}\right]=0,\nonumber
\eqn
where $\gamma(a):=\frac{2^9}{3^2}a(3^4-225a+276a^2-154a^3+32a^4)$.
Considering the analogous scenario on the other quartets $13|24$ and $14|23$, we have 
\beqn
\gamma(a)-E\left[\frac{q_1}{\det(P_{13})\det(P_{24})}\right]=0,\qquad 
\gamma(a)-E\left[\frac{q_2}{\det(P_{14})\det(P_{23})}\right]=0.\nonumber
\eqn 
respectively.
 
We can use the observed pattern frequencies $\bf{f}$ to estimate the probability distributions $P_{12}$ and $P_{34}$, and either, assuming $a$ is small, a quadratic approximation or a general root finding algorithm to find the smallest positive root of the above equation.
In this way, we obtain an estimate $\widehat{a}$ of the internal edge length of a quartet tree under the assumption of JC on the internal edge and GM on the pendant edges.

We see that the squangle method is capable of returning two types of quartet weights.
For each quartet, the first gives a confidence value, and the second gives a measure of genetic distance along the internal edge.

Below we compare these weights to those computed from maximum-likelihood inference, where, as is described in \citet{StRa2002}, weights are calculated as
\[w_i = \frac{L_i}{L_1 + L_2 + L_3},\]
where $L_i$ is the likelihood of quartet $i$.

\subsection*{Simulation study}

\begin{table}[tb]
\begin{center}
\begin{tabular}{lllll}
 & A & C & G & T \\
A & $1-a$      & $ab/(2b+1)$ & $ab/(2b+1)$ & $a/(2b+1)$ \\
C & $a/(b+2)$  & $1-a$       & $ab/(b+2)$  & $a/(b+2)$ \\
G & $a/(b+2)$  & $ab/(b+2)$  & $1-a$       & $a/(b+2)$ \\
T & $a/(2b+1)$ & $ab/(2b+1)$ & $ab/(2b+1)$ & $1-a$ \\ 
\end{tabular}  
\caption{Structure of the substitution matrices used in the simulations. $b\geq 1$ parametrises GC bias.} 
\label{tab:submat}
\end{center} 
\end{table}

\begin{figure}[ht]
\centering
  \subfloat[]{\label{fig:2a}\includegraphics[width=.6\textwidth]{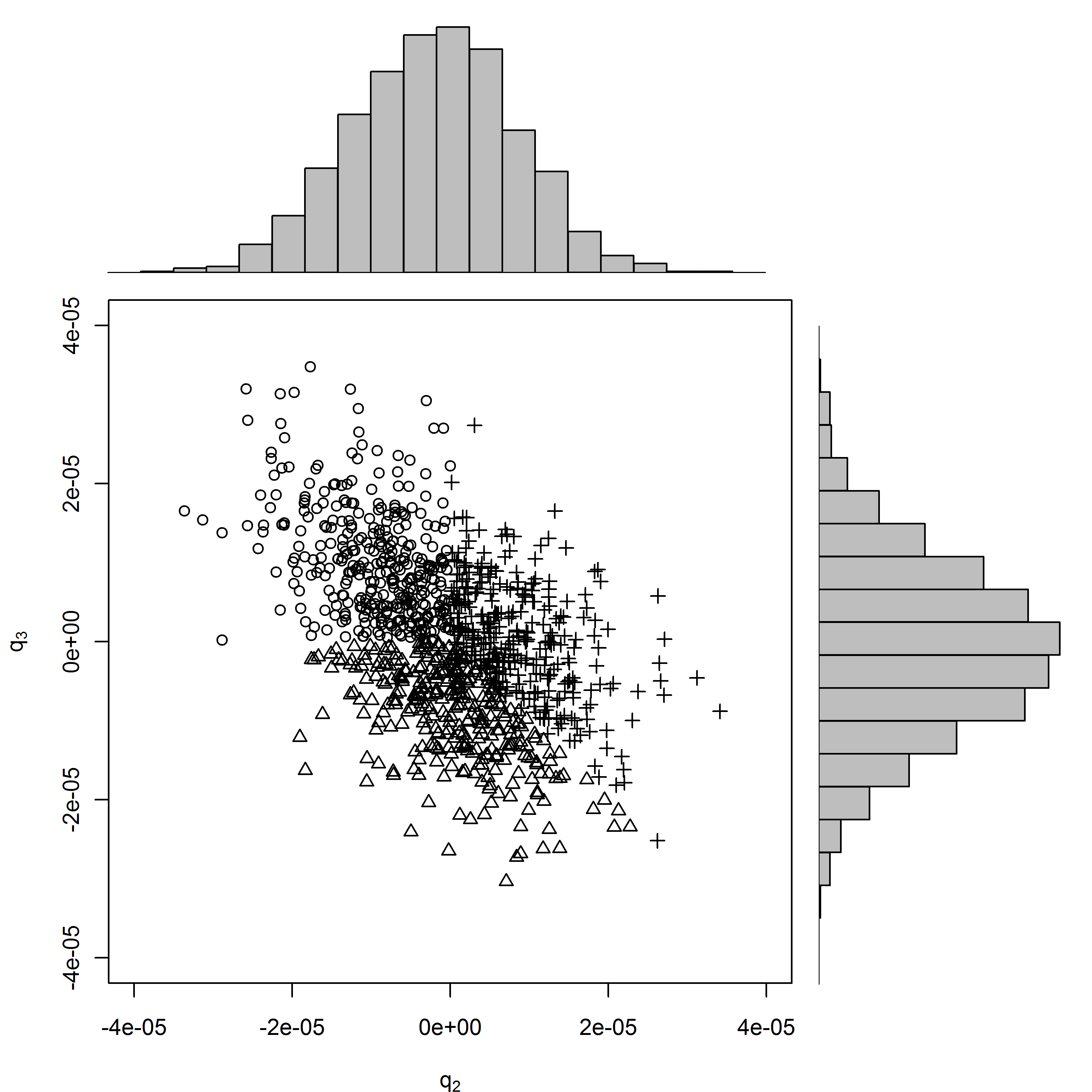}}
\\
  ~ 
  \subfloat[]{\label{fig:2b}\includegraphics[width=.6\textwidth]{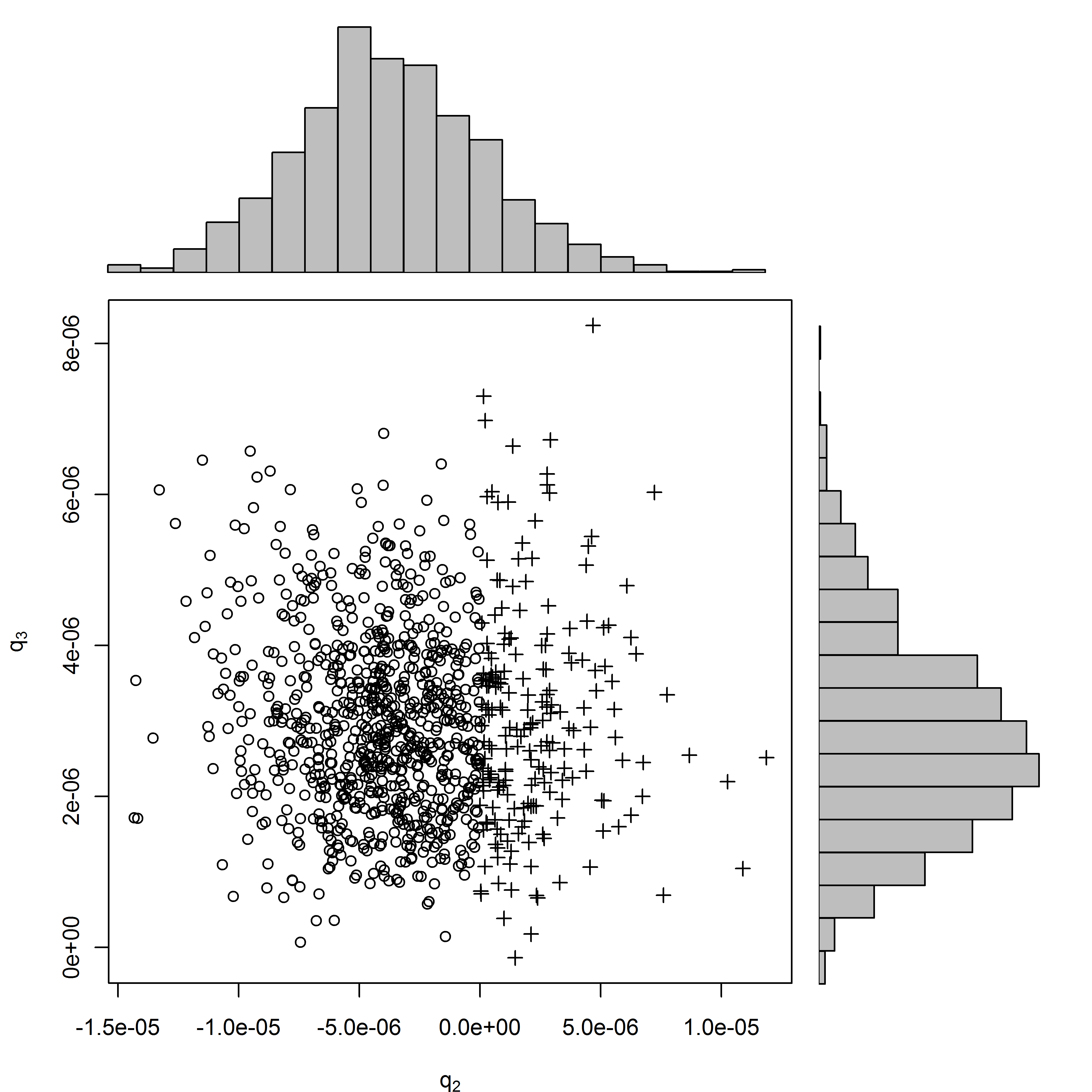}}
\caption{The distribution of $q_2$ and $q_3$ for data simulated on (a) the star tree, and (b) the Felsenstein tree. 
In each case, the quartet returned by the least-squares routine is shown by different plotting symbols: $\circ$ $12\mid34$, $\triangle$ $13\mid24$ and $+$ $14\mid23$.
\label{fig:q2q3dists}}
\end{figure}

We conducted two simulations to assess the assumption that, taken in pairs, the squangles are independent and normally distributed about their respective means with identical standard deviation. 
We simulated data on a star tree under the JC model, where $M$, the substitution matrix for each pendant edge, had 0.9 for diagonal entries (i.e. in terms of Table~\ref{tab:submat}, $a=0.1$ and $b=1$).
We next simulated data on the quartet $12|34$ where $M_s$, the substitution matrix for each short edge, had  $a=0.02$ and $b=1$, and $M_l$, the substitution matrix for each long edge, had $a=0.4$ and $b=5$.
We recorded all the values of $q_2$ and $q_3$ along with the tree topology selected and plotted these values on the $q_2$-$q_3$ plane (as shown in Figure~\ref{fig:q2q3dists}) along with their marginal distributions.

We tested the strength of the relationship (\ref{eq:intedgeval}) between the squangle values and internal edge lengths  for finite data.
This was done by conducting simulations with sequence length set at 1000 sites under JC -- so all edges had $b=1$ (i.e. no GC bias).
On the pendant edges we set $a=0.1$ and on the internal edge $a$ was taken in the range 0.01 to 0.20. 
For each value of $a$, we performed 1000 repetitions. 
 
Our primary simulations tested the robustness of the least-squares implementation of the squangle method, and were conducted on 
(i) the ``Standard'' tree, which has has one short internal edge and 4 long pendant edges of equal length, 
(ii) the ``Felsenstein'' tree, and (iii) the ``Farris'' tree.
The Felsenstein and Farris trees each have a combination of 3 short and 2 long (external) edges: on the Felsenstein tree, the long edges are two non-sister external edges, and on the Farris tree they are two sister external edges.
Parameters varied were: 
\begin{enumerate}
\item Sequence length $c \in \{500, 1000\}$; 
\item GC bias $b \in \{1,3,5,7,9\}$;
\item Short edge: $a \in \{0.01, 0.02,0.04\}$; 
\item Long edge: $a \in \{0.2, 0.4\}$;
\item Proportion of invariant sites $\nu \in \{0,0.25,0.5,0.75\}$
\end{enumerate}
Each parameter setting was run 1000 times and trees were estimated using: the least-squares implementation of the squangles \emph{SQ}; least-squares squangles modified to down-weight the observed frequency of constant sites \emph{SQi}; neighbor-joining \emph{NJ}, where distances were estimated under the JC model; \emph{NJ} using logDet distances \emph{NJld}; maximum-likelihood under the JC model \emph{ML}; maximum-likelihood under the JC model plus invariant sites \emph{MLi}; and maximum parsimony \emph{MP}.

\subsection*{Supertrees or supernetworks}

Most phylogenetic data sets of interest contain more than four taxa, so to perform inference using the squangles it is necessary to break the data into groups of four taxa, infer a quartet for each group of four taxa, and then piece these quartet trees back together into an overall tree (or perhaps network).
We have produced \emph{Python} code that reads a sequence alignment in \emph{Phylip} \citep{phylip} format and uses the squangles to calculate a set of weighted quartets. 
The squangles cannot deal with data that includes gaps or ambiguous characters, so there is an option to either remove such sites at the beginning of the analysis from the whole alignment (faster), or on a quartet by quartet basis (slower but retains more information).
There is also an option to account for invariant sites by implementing \emph{SQi}; this uses the method of \citet{St2000} and hence must be done on a quartet by quartet basis. 
Options for output format include:
\begin{itemize}
\item A simple list of the best tree for each quartet (this could be used as input to \emph{MRP}, for instance using \emph{Clann} \citep{CrMc2005}, or could be opened in \emph{SplitsTree4} \citep{HuBr2006} to 
create a $Z$-closure network); 
\item A  list of each quartet with confidence higher than some threshold value (this could be used as above);
\item A file in QuartetSuite format including three confidence weights for each set of 4 taxa;
\item A file in QNet format that includes three distance weights for each set of 4 taxa.
\end{itemize}
Python code is available from BRH at \emph{barbara.holland@utas.edu.au}.

\subsection*{Case studies}

We applied the squangle method to two biological data sets that have been discussed in the context of compositional heterogeneity.

\subsubsection*{Mammal data}

\citet{PhPe2003} used a data set consisting of mtDNA for mammals (and outgroups) to test evidence for Theria (a hypothesis which groups Marsupials and Placentals) against Marsupionta (a hypothesis that groups Marsupials and Monotremes). 
The data set contains two monotremes (Platypus and Echidna),  nine placental mammals, four marsupials (Opossum, Bandicoot, Brushtail, Wallaroo) and ten outgroup taxa (birds and reptiles).
\citet{PhPe2003} found RY coding of 3rd codon positions decreased support for Marsupionta in comparison to Theria.
Since Theria is more frequently supported by analyses of nuclear genes, they argued that the support for Marsupionta could be an artefact of base-composition bias.
We used the squangle method to evaluate support for 3 competing hypotheses:\newline

\begin{tabular}{ll}
\textbf{Marsupionta} & (Outgroup, ((Monotreme, Marsupial), Placental));\\
\textbf{Theria} & (Outgroup, ((Placental, Marsupial), Monotreme));\\
\textbf{Other}  & (Outgroup, ((Placental, Monotreme), Marsupial)).\\
\end{tabular}
\newline

We investigated if squangles appeared more robust to the apparent base-composition bias than a standard likelihood approach.
We compared \emph{ML} trees to squangle-based trees for the $2\cdot 10\cdot 4\cdot 9=720$ quartets that include one taxon from each of the four groups.
Following \citet{PhPe2003}, \emph{ML} was performed using the TN93 model \citep{tamura1993} with and without allowing for invariant sites.
We used the squangles with and without the invariant sites corrections, assigned confidence-style weights to each quartet and analysed each codon position separately.
This allowed us to test if high-confidence quartets were more likely to support any particular hypothesis.

Additionally, we ran three analyses -- one for each codon position -- on all of the ${25 \choose 4}  = 12,650 $ subsets of four taxa.
Quartets with confidence weight greater than 0.95 were retained and used as input to the \emph{MRP} supertree method.

\subsubsection*{Possum data}

\citet{Gr2007} investigated the phylogenetic relationships for 41 species of possums. 
They found that the gene RAG gave very different results to all the other genes previously investigated and to morphological data. 
With reference to their Figure~1, all the data sets they analysed gave a clade they label B (which contains 10 species from the genera {\em Gracilinanus}, {\em Cryptonanus}, and {\em Thylamys}), and a clade they label I (which contains 8 species from the genera {\em Micoureus} and {\em Marmosa}). 
For all the data sets apart from RAG they found that clade B was sister to the {\em Marmosops} genus (5 species), with clade I grouping elsewhere in the tree. 
However, for the RAG gene clade B and clade I were sister to each other.
\citet{Gr2007} concluded that the grouping of clade B with clade I for the RAG gene was the result of a base-composition artefact.
Using the same approach as described for the \citet{PhPe2003} data, we evaluated support for 3 competing hypotheses
for each of the $10\cdot 8\cdot 5\cdot 18=7200$ quartets that include one taxon from each of the 4 groups:
\newline

\begin{tabular}{ll}
\textbf{True} &  ((B, \emph{Marmosops}), I, Remainder); \\
\textbf{GC-bias} & ((B, I), \emph{Marmosops}, Remainder);\\
\textbf{Other} & ((B, Remainder), \emph{Marmosops}, I).
\end{tabular} 
\newline

Likelihoods were calculated under the GTR model both with and without invariant sites. 
Again, we investigated if high-confidence quartets were more likely to support the presumed true tree. 
For the \citet{Gr2007} data set we also investigated 1st and 2nd codon positions combined, and all three codon positions combined.

For the RAG gene and the non-RAG genes we also ran three analyses each -- one for each codon position -- of all ${41 \choose 4}=101,270$ quartets. 
Quartets with confidence weight greater than 0.95 were retained and used as input to the \emph{MRP} supertree method.

\section*{Results}

\subsection*{Simulation study}

Figures~\ref{fig:q2q3dists}(a) and (b) show results of 1000 simulations with sequence lengths of 1000 sites. 
The values of the squangles $q_2$ and $q_3$ are plotted in the plane and the marginal distribution for each squangle is shown as a histogram.
Figure~\ref{fig:q2q3dists}(a) presents data simulated on a star tree under the JC model (ie. $b=1$), where $M$, the substitution matrix for each pendant edge had $a=0.1$.
The correlation between $q_2$ and $q_3$ was $r=-0.519$.
Figure \ref{fig:q2q3dists}(b) is for data simulated on the quartet $12|34$ for the a Felsenstein  tree.
$M_s$, the substitution matrix for each short edge, had  $a=0.02$ and $b=1$, and $M_l$, the substitution matrix for each long edge, had  $a=0.4$ and $b=5$.
The correlation between $q_2$ and $q_3$ was $r=-0.053$.

We conclude that the distribution of both $q_2$ and $q_3$ seems to be approximately normal, but with some skew (particularly for the Felsenstein-tree simulation) and some violation of independence (particularly for the star-tree simulation).

\begin{figure}[ht]
\centering
\includegraphics[width=.9\textwidth]{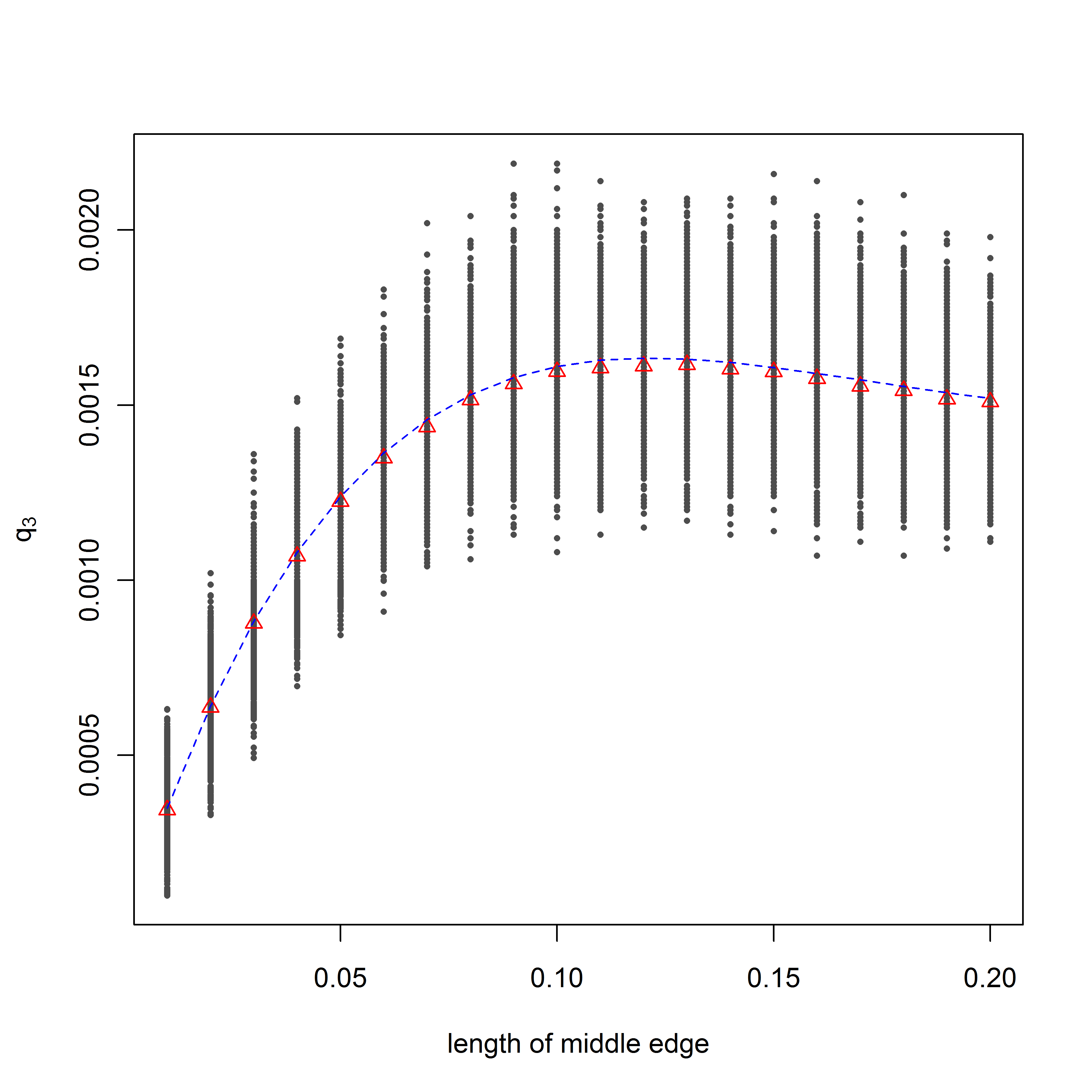}
\caption{$q_3$ against internal edge length $a$.
The blue dashed line shows the theoretical value according to equation (\ref{eq:intedgeval}), the dots show each simulation, the red triangles show the mean value for the simulations at that value of $a$.
\label{fig:midedge}}
\end{figure}

Figure~\ref{fig:midedge} shows how $q_3$ varies with internal edge length compared to the theoretical expectation under the JC model.
At sequence length $c=1000$ there is good agreement between the mean of the 1000 simulations for each value of $a$ and the theoretical expectation.

\begin{figure}[ht]
\centering
\includegraphics[width=36em]{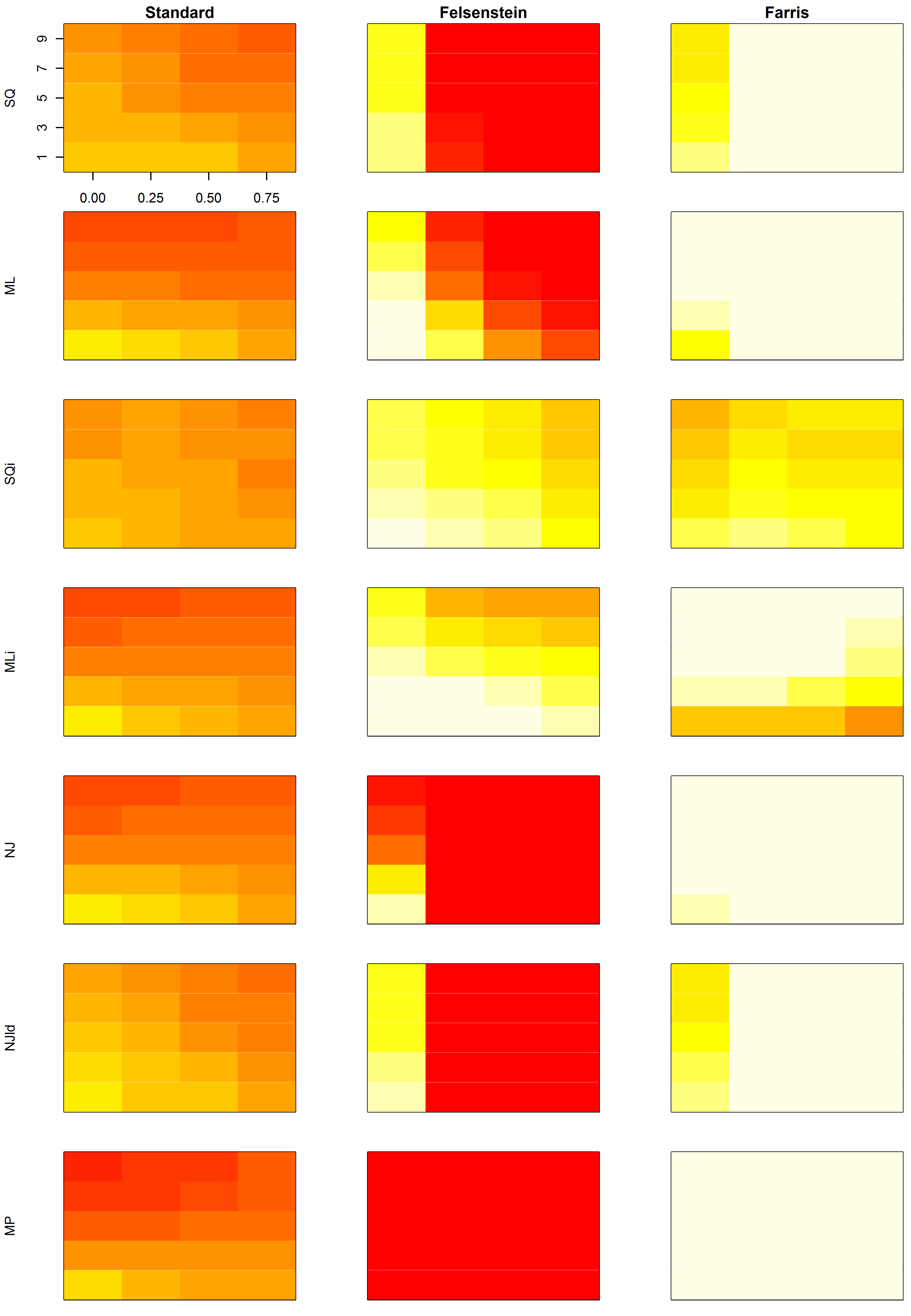}
\caption{Main simulation results.
Each rectangle corresponds to the accuracy of a particular method (one method per row), on a particular tree type (one tree type per column).
Within the rectangle the $x$-axis relates to the proportion of invariant sites increasing from left to right; and the $y$-axis relates to the GC bias, increasing from bottom to top. 
Hence, the bottom left corner of each rectangle corresponds to the most straight-forward case for phylogenetic inference. 
The darkest red shade (found for \emph{MP} on the Felsenstein tree) corresponds to 0\% accuracy, and the lightest shade (found for MP on the Farris tree) corresponds to 100\% accuracy.
Simulations settings shown here are $c=1000$, short edge $a=0.02$, long edge $a=0.4$.
\label{fig:mainsim}}
\end{figure}

Figure~\ref{fig:mainsim} shows the results for short edge $a = 0.02$, long edge $a = 0.4$, and sequence length $c=1000$.
The \emph{SQ} method suffers markedly from long branch attraction on the Felsenstein tree whenever invariant sites are present. 
\emph{SQi} performs much better than \emph{SQ} when invariant sites are present, and there is still a slight decrease in accuracy as the proportion of invariant sites increases; however this is similar to what occurs for \emph{MLi}.
On the Farris tree all the methods seem biased towards getting the tree correct; although in this respect \emph{SQi} seems to be relatively less positively biased.
On the Standard  and Felsenstein trees, there are regions where both \emph{SQ} and \emph{SQi} are more accurate than \emph{ML} or \emph{MLi}.
This occurs when the proportion of invariant sites is low and the base-composition bias is high, although, as has been previously noted in \citet{GaGo1995}, \emph{ML} is reasonably robust to violations of its model assumptions.
\emph{SQ} has similar accuracy to \emph{NJ} with logDet distances.

\subsection*{Case studies}

\subsubsection*{Mammal data}

\begin{table}[tb]
\tiny
  \centering
  \caption{Analysis of 720 quartets from the \citet{PhPe2003} data set by codon position. Numbers of each of the three possible quartet topologies are shown with high-confidence quartets ($\geq 0.95$ confidence weight for \emph{SQ} and \emph{SQi}, and $\geq 0.95$ likelihood weight for \emph{ML} and \emph{MLi}) shown in brackets.}
    \begin{tabular}{r|rrr|rrr|rrr}
    \toprule
    position & 1st   &       &       & 2nd   &       &       & 3rd   &       &  \\
    \midrule
    method & Marsupionta & Other & Theria & Marsupionta & Other & Theria & Marsupionta & Other & Theria \\
    \emph{ML} & 606 (553) & 47 (26) & 67 (42) & 477 (395) & 210 (159) & 33 (15) & 322 (174) & 184 (84) & 214 (103) \\
    \emph{MLi} & 530 (301) & 85 (16) & 105 (30) & 273 (87) & 372 (85) & 75 (6) & 310 (159) & 195 (88) & 215 (104) \\
    \emph{SQ}    & 604 (190) & 57 (10) & 59 (7) & 549 (172) & 157 (10) & 14 (2) & 268 (54) & 223 (33) & 229 (32) \\
    \emph{SQi}   & 445 (120) & 166 (27) & 109 (16) & 321 (74) & 351 (99) & 48 (9) & 235 (50) & 234 (39) & 251 (43) \\
    \bottomrule
    \end{tabular}
  \label{tab:Phillips}
\end{table}

For the \citet{PhPe2003} data sets, Table \ref{tab:Phillips} shows the performance of \emph{SQ} and \emph{SQi} compared to \emph{ML} using a TN93 model both with and without invariants sites.
Overall this data set produces far fewer high-confidence quartets than the \citet{Gr2007} data set.
The \emph{SQi} method with partitioning by codon position makes little headway in recovering the suggested true tree, with Marsupionta strongly preferred over Theria for 1st and 2nd codon positions.  
Third codon positions slightly favour Theria.
\emph{MLi} does little better than \emph{SQi} at picking the Theria topology.
Considering all quartets, \emph{SQi} is slightly better for codon position 2, but worse for positions 1 and 3.
Considering only high-confidence quartets, \emph{SQi} is better for positions 1 and 3 but slightly worse for position 2.  

\begin{figure}[htb]
\centering
\includegraphics[width=1.5\textwidth]{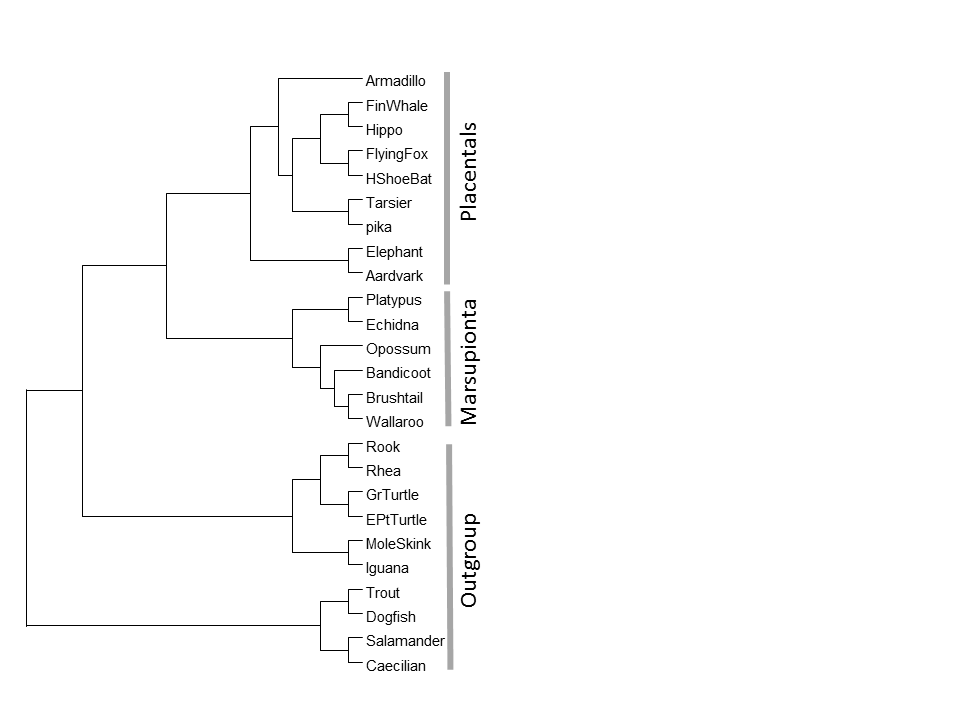}
\caption{\emph{MRP} supertree of all high-confidence quartets found by the \emph{SQi} method applied to 1st,
2nd and 3rd codon positions of the \citet{PhPe2003} data set.
\label{fig:phillipsMRP}}
\end{figure}

The \emph{MRP} tree constructed from high-confidence \emph{SQi} quartets from each codon position (see Figure \ref{fig:phillipsMRP}), supports the Marsupionta hypothesis.
Apart from this contentious point, it is congruent with previous studies.

\subsubsection*{Possum data}

\begin{table}[tb]
\tiny
  \centering
  \caption{Analysis of 7200 quartets from the RAG gene of the \citet{Gr2007} data set by codon position. Numbers of each of the three possible quartet topologies are shown with
  high-confidence quartets ($\geq 0.95$ confidence weight for \emph{SQ} and \emph{SQi}, and $\geq 0.95$ likelihood weight for \emph{ML} and \emph{ML}i) shown in brackets.}
    \begin{tabular}{r|rrr|rrr|rrr}
    \toprule
    position & 1st   &       &       & 2nd   &       &       & 3rd   &       &  \\
    \midrule
    method & GC-bias & Other & True & GC-bias & Other & True & GC-bias & Other & True \\
    \emph{ML}    & 3538 (2203) & 667 (228) & 2995 (1070) & 4819 (2418) & 2279 (814) & 102 (0) & 7174 (7132) & 0 (0) & 26 (7) \\
    \emph{MLi}   & 3206 (493) & 965 (132) & 3029 (72) & 4819 (517) & 2244 (295) & 137 (0) & 7069 (6863) & 7 (0) & 124 (29) \\
    \emph{SQ}    & 3334 (464) & 680 (80) & 3186 (731) & 4549 (1152) & 2418 (222) & 233 (55) & 7181 (5274) & 0 (0) & 19 (0) \\
    \emph{SQi}   & 1661 (197) & 918 (168) & 4621 (1149) & 4186 (875) & 2701 (386) & 313 (80) & 5795 (3044) & 294 (56) & 1111 (264) \\
    \bottomrule
    \end{tabular}
  \label{tab:RAG}
\end{table}

For the RAG gene of the \citet{Gr2007} data set, Table \ref{tab:RAG} shows the performance of \emph{SQ} and \emph{SQi} compared to \emph{ML} using a GTR model with and without allowing for invariant sites.
None of the methods do particularly well: they more often choose the ``GC-bias" tree than the true tree for all partitions, with the exception of \emph{SQi} for codon position 1. 
When comparing all 7200 quartets, \emph{SQi} has more instances of the ``true" tree and fewer instances of the  ``GC-bias" tree than \emph{MLi} for all three codon positions. 
Restricting the comparison to high-confidence quartets, \emph{SQi} picks more instances of the ``true" tree  than \emph{MLi} for all codon positions, 
but for the 2nd codon position it also picks more of the ``GC-bias" tree. 
For both \emph{ML} and \emph{SQ} adding invariant sites  to the model reduces the number of incorrect trees chosen; 
accounting for invariant sites also reduces the number of high-confidence quartets overall.
 Results for combinations of partitions of the RAG gene are poor compared to analyses of single partitions (see Table~\ref{tab:RAGcombo}).
All methods suffer a severe decline in performance for data that combines all three codon positions.

\begin{table}[tb]
\tiny
  \centering
  \caption{Analysis of 7200 quartets from the \citet{Gr2007} data set for a concatenation of 1st and 2nd codon positions and for all codon positions. 
  Numbers of each of the three possible quartet topologies are shown with
  high-confidence quartets ($\geq 0.95$ confidence weight for \emph{SQ} and \emph{SQi}, and $\geq 0.95$ likelihood weight for \emph{ML} and \emph{MLi}) shown in brackets.}
    \begin{tabular}{r|rrr|rrr}
    \toprule
    position & \multicolumn{3}{c}{1st and 2nd combined} & \multicolumn{3}{c}{1st, 2nd and 3rd combined} \\
    \midrule
    method & GC-bias & Other & True & GC-bias  & Other & True \\
   \emph{ML}    & 4978 (3823) & 1112 (596) & 1110 (376) & 7197 (7196) & 0 (0) & 3 (0) \\
    \emph{MLi}   & 4923 (1292) & 1192 (276) & 1085 (22) & 7186 (7100) & 10 (3) & 4 (0) \\
    \emph{SQ}    & 4756 (1833) & 1140 (215) & 1304 (272) & 7198 (5653) & 1 (0) & 1 (0) \\
    \emph{SQi}   & 3378 (1099) & 1479 (258) & 2343 (493) & 6926 (4236) & 170 (7) & 104 (3) \\
    \bottomrule
    \end{tabular}
  \label{tab:RAGcombo}
\end{table}

\begin{table}[tb]
\tiny
  \centering
  \caption{Analysis of 7200 quartets from the non-RAG genes of the \citet{Gr2007} data set by codon position. Numbers of each of the three possible quartet topologies are shown with
  high-confidence quartets ($\geq 0.95$ confidence weight for \emph{SQ} and \emph{SQi}, and $\geq 0.95$ likelihood weight for \emph{ML} and \emph{MLi}) shown in brackets.}
    \begin{tabular}{rrrrrrrrrr}
    \toprule
    position & \multicolumn{3}{c}{1st} & \multicolumn{3}{c}{2nd} & \multicolumn{3}{c}{3rd} \\
    \midrule
    method & GC-bias & Other & True & GC-bias  & Other & True & GC-bias  & Other & True \\
   \emph{ ML}    & 226 (20) & 179 (4) & 6795 (5026) & 1455 (342) & 1172 (79) & 4573 (2027) & 88 (0) & 618 (169) & 6494 (4894) \\
    \emph{Mli}   & 254 (0) & 341 (0) & 6605 (1418) & 1589 (57) & 1337 (8) & 4274 (1254) & 77 (0) & 657 (25) & 6466 (1611) \\
    \emph{SQ}    & 236 (0) & 67 (3) & 6897 (1721) & 557 (62) & 642 (42) & 6001 (1871) & 129 (2) & 463 (37) & 6608 (3262) \\
    \emph{SQi}   & 291 (8) & 81 (1) & 6828 (1637) & 446 (40) & 609 (32) & 6145 (1880) & 161 (15) & 528 (57) & 6511 (3330) \\
    \bottomrule
    \end{tabular}
  \label{tab:nonRAG}
\end{table}

For the non-RAG genes of the \citet{Gr2007} data set, Table \ref{tab:nonRAG} shows the performance of \emph{SQ} and \emph{SQi} compared to \emph{ML} using a GTR model with and without allowing for invariant sites.
When comparing all 7200 quartets \emph{SQi}, has more instances of the true tree and fewer instances of the bias tree both than \emph{ML} and \emph{MLi}; and this is true for all three codon positions. 
Intriguingly, for this example, adding invariant sites to the \emph{ML} model increases the number of incorrect topologies chosen.
Restricting the comparison to high-confidence quartets \emph{SQi}, picks more true trees than \emph{MLi} for all codon positions, but fewer true trees than \emph{ML} without invariant sites. 

\begin{figure}[tb]
\centering
\includegraphics[width=1.7\textwidth]{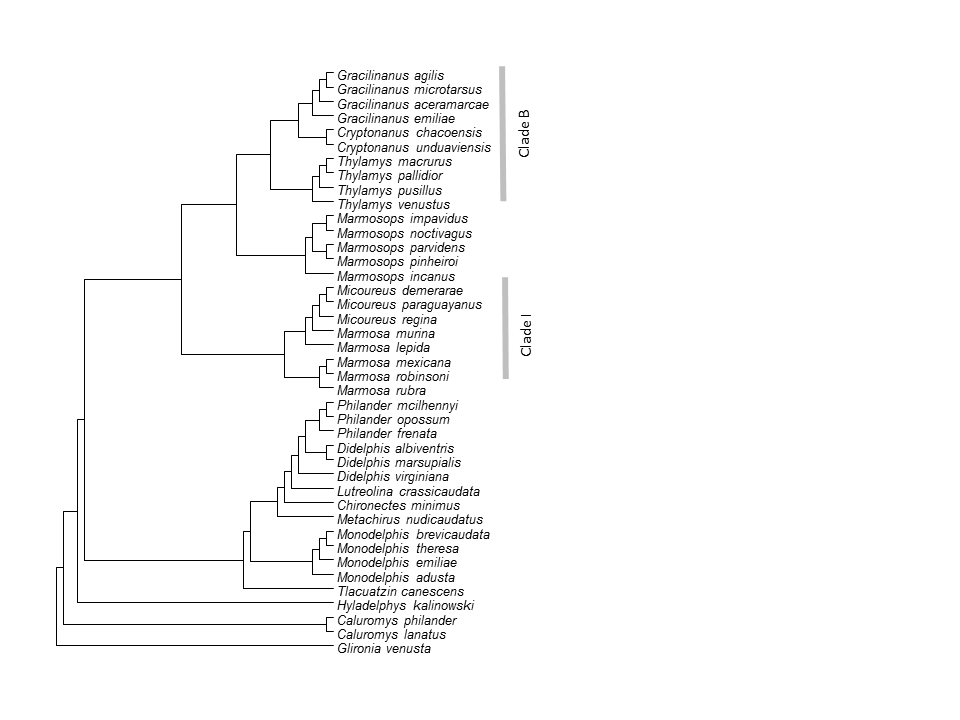}
\caption{\emph{MRP} supertree of all high-confidence quartets found by the \emph{SQi} method applied to 1st,
2nd and 3rd codon positions of the RAG gene and the 1st, 2nd and 3rd codon positions of the non-RAG genes from the \citet{Gr2007} data set.
\label{fig:gruberMRP}}
\end{figure}

The \emph{MRP} supertree formed from only high-confidence quartets from each codon position in both the RAG and non-RAG genes is shown in Figure~\ref{fig:gruberMRP}.
It recovers the presumed true relationship between the {\em Marmosops} genus and Clade B; the rest of the tree is also broadly congruent with the findings of \citet{Gr2007}.
This is a very pleasing result as \citet{Gr2007} found that analysis of this combined  data under a variety of phylogenetic methods produced the artefactual grouping of clade B with clade I.
However, if only high-confidence quartets from codon partitions of the RAG gene are considered, then the \emph{MRP} tree does include the artefactual B+I grouping (results not shown).

\subsubsection*{Efficiency}

While running the \emph{SQ} and \emph{SQi} methods on the case study data we timed some of the runs in order to check the efficiency of the python implementation of the method. 
For codon position 1 of the \citet{PhPe2003} data (3588 base pairs), the analysis of the 720 quartets for Table \ref{tab:Phillips} took 90 seconds (approximately 0.125 seconds per quartet).
For the combined data (10764 base pairs) the analysis of the 720 quartets took 98 seconds.

\section*{Conclusions}

The squangles were originally presented in \citet{Su2008,SuJa2009}, here we provide the first investigation of how robust the squangle method is to violation of the IID assumption.
We found that \emph{SQ} is fairly robust to the presence of invariant sites for the standard (clock-like) simulation topology.
 However, with the Felsenstein topology performance degraded with only a small proportion of invariant sites. 
Encouragingly, a simple modification of the squangle method that estimates and removes invariant sites greatly improves overall accuracy.
For data generated under a process that incorporates both a proportion of invariant sites and base-composition drift, the newly proposed  \emph{SQi} method is comparable to, or better than, maximum-likelihood under a homogeneous invariant sites model.
Results from the \citet{Gr2007} case study also indicate that \emph{SQi} is more robust to changing base-composition than \emph{ML}.

We also introduced two new weighting schemes for quartets derived using the squangle method: one based on confidence in the quartet and other based on length of the internal edge.
Restricting to high-confidence quartets improved accuracy in both the case studies. 
Combining high-confidence quartets  from different genes and codon positions gave the presumed correct tree for the \citet{Gr2007} case study; whereas other methods of analysing this data are typically misled by the strong base-composition bias in the RAG gene.
It was disappointing that the approach of combining high-confidence quartets from different codon positions did not recover the ``true" tree in either the \citet{PhPe2003} case study, or the \citet{Gr2007} case study when the data was restricted to just the RAG gene.
Perhaps it was too hopeful to think that accounting for base-composition shift and invariant sites would be enough for unbiased phylogenetic estimation in these cases. 
We recall that while \emph{SQi} can account for data that evolved under GM with invariant sites it is not expected to be consistent if there is rate-variation across sites.
\emph{SQi} will also not be consistent if there are mixtures of processes, e.g. as would be expected if different groups of sites were evolving under different models \citep{PaMe2004, KoTh2004}. 
The results for the case studies suggest that some of these extra unmodelled factors are probably at play.
The case studies support one's intuition that partitioning by codon position should mitigate against the problem of mixtures of processes.

To revisit the five questions posed at the end of the introduction:
\begin{enumerate}
\item  We have shown that viewing the squangles in a least-squares framework gives a natural way of assigning confidence-based weights to quartets.
Furthermore, the case studies suggest that restricting to high-confidence quartets improves the ratio of correct to incorrect topologies;
\item We illustrated the squangle method for tree-building by using them in conjunction with \emph{MRP}, however there are many options for utilising quartet information to build trees or supertrees, and the output of the \emph{SQi} method could be used as input to any of these;
\item We have shown that there is a region of parameter space in which squangles, particularly with invariant sites accounted for, 
provide more accurate phylogenetic reconstruction than maximum-likelihood.
As expected given the different assumptions of the \emph{SQi} and \emph{MLi} methods, \emph{SQi} does comparatively better when there is base-composition bias, whereas \emph{MLi} does better on data without base-composition bias;
\item We found that unmodified squangles were not particularly robust to the presence of invariant sites, particularly for the Felsenstein tree-shape;
\item However, this could be remedied by applying the method of \citet{St2000} to estimate and subsequently remove a proportion of invariant sites.
\end{enumerate}

Overall our results show that \emph{SQi} provides a complementary tool to existing methods in the phylogenetic toolkit.
We suggest that the \emph{SQi} method should be used alongside maximum-likelihood approaches in cases where base-composition differs across the taxa under consideration.

\section*{Acknowledgements}

The authors would like to thank Matt Phillips for supplying the mammal data from \citet{PhPe2003}.

\section*{Funding}
This research was conducted with support from Australian Research Council (ARC) Future Fellow Grant FT100100031 (BRH), and ARC Discovery Grant DP0877447 (JGS and PDJ).

\section*{Appendix}

Here we derive the value of the squangle $q_1$ when evaluated on joint probability distribution $P$ arising from the quartet $13|24$ with GM substitution matrices $M_i$ on the pendant edges and a JC substitution matrix with off diagonal entry $\frac{a}{3}$ on the internal edge.
The entries of $P$ are represented as an array where $P_{AGGC}$ is the probability of observing nucleotide states $A,G,G$ and $C$ at leaf 1 through to 4 respectively.
We start by writing $q_1$ as a sum of two polynomials $q_1=f^{(13,24)}-f^{(14,23)}$, and recall the expressions given in \citet{SuJa2009}:
\beqn\label{eq:squangparts}
f^{(13,24)}(P)&=\lambda\sum_{j_1,k_1,\ell_1,m_1,i_2,k_2,\ell_2,m_2\in \{A,C,G,T\}} \widehat{P}_{\Sigma i_2}\widehat{P}_{j_1\Sigma}\widehat{P}_{k_1k_2}\widehat{P}_{\ell_1\ell_2}\widehat{P}_{m_1m_2}\epsilon_{j_1k_1\ell_1m_1}|\epsilon_{i_2k_2\ell_2m_2}|,\\
f^{(14,23)}(P)&=\lambda\sum_{j_1,k_1,\ell_1,m_1,i_2,k_2,\ell_2,m_2\in \{A,C,G,T\}} \widehat{P}_{j_1i_2}^2\widehat{P}_{k_1k_2}\widehat{P}_{\ell_1\ell_2}\widehat{P}_{m_1m_2}|\epsilon_{j_1k_1\ell_1m_1}|\epsilon_{i_2k_2\ell_2m_2}|,
\eqn
where $\lambda=\prod_{i=1}^4\det(M_i)$; $\widehat{P}(AT)$ is the probability of observing the states $A,T$ at the internal vertices of the quartet; $\widehat{P}({\Sigma T})=\sum_{i\in \{A,C,G,T\}}\widehat{P}(iT)$; and $|\epsilon_{ijk\ell}|=0$ if any of $i,j,k$ and $\ell$ are equal, and is 1 otherwise.

Under the JC assumption for the internal edge, we have $\widehat{P}_{ij}=\frac{1}{4}\left(\frac{a}{3}+\delta_{ij}(1-\frac{4a}{3})\right)$; where $\delta_{ij}=1$ if $i=j$ and 0 otherwise, and $\widehat{P}_{\Sigma i}=\widehat{P}_{i\Sigma}=\frac{1}{4}$ for all $i\in \{A,C,G,T\}$.
We also notice that we can use the properties of $|\epsilon_{ijk\ell}|$ to simplify the summations (\ref{eq:squangparts}).
For instance, given any array $X_{ijk\ell}$, we have
\beqn\label{eq:simpsum}
\sum_{i,j,k,\ell\in\{A,C,G,T\}}X_{ijk\ell}|\epsilon_{ijk\ell}|=\sum_{\sigma\in S}X_{\sigma(A)\sigma(C)\sigma(G)\sigma(T)},
\eqn
where $S$ is the set of permutations on the letters $A,C,G$ and $T$.
Applying the idea that underlies (\ref{eq:simpsum}) twice, we have

\beqn
f^{(13,24)}(P)&=\lambda\sum_{\sigma_1,\sigma_2\in S}\widehat{P}_{\Sigma \sigma_2(A)}\widehat{P}_{\sigma_1(A)\Sigma }\widehat{P}_{\sigma_1(C)\sigma_2(C)}\widehat{P}_{\sigma_1(G)\sigma_2(G)}\widehat{P}_{\sigma_1(T)\sigma_2(T)}\nonumber\\
&=\frac{\lambda}{4^5}\sum_{\sigma_1,\sigma_2\in S}\left(\fra{a}{3}+\delta_{\sigma_1(C)\sigma_2(C)}(1\!-\!\fra{4a}{3})\right)\left(\fra{a}{3}+\delta_{\sigma_1(G)\sigma_2(G)}(1\!-\!\fra{4a}{3})\right)\left(\fra{a}{3}+\delta_{\sigma_1(T)\sigma_2(T)}(1\!-\!\fra{4a}{3})\right),
\eqn
and
\beqn
f^{(14,24)}(P)&={\lambda}\sum_{\sigma_1,\sigma_2\in S}\widehat{P}_{\sigma_{1}(A)\sigma_2(A)}^2\widehat{P}_{\sigma_1(C)\sigma_2(C)}\widehat{P}_{\sigma_1(G)\sigma_2(G)}\widehat{P}_{\sigma_1(T)\sigma_2(T)}\\
&=\frac{\lambda}{4^5}\sum_{\sigma_1,\sigma_2\in S}\left(\fra{a}{3}+\delta_{\sigma_1(A)\sigma_2(A)}(1\!-\!\fra{4a}{3})\right)^2\left(\fra{a}{3}+\delta_{\sigma_1(C)\sigma_2(C)}(1\!-\!\fra{4a}{3})\right)\left(\fra{a}{3}+\delta_{\sigma_1(G)\sigma_2(G)}(1\!-\!\fra{4a}{3})\right)\\
&\hspace{20em}\cdot\left(\fra{a}{3}+\delta_{\sigma_1(T)\sigma_2(T)}(1\!-\!\fra{4a}{3})\right).\nonumber
\eqn

From here the problem is purely combinatorial.
For instance, it is not difficult to show
\beqn
&\sum_{\sigma_1,\sigma_2\in S}1= 4! \cdot 4! = 24 \cdot 24,\\
&\sum_{\sigma_1,\sigma_2\in S}\delta_{\sigma_1(A)\sigma_2(A)}= 4! \cdot 3! =24\cdot 6,\\
&\sum_{\sigma_1,\sigma_2\in S}\delta_{\sigma_1(A)\sigma_2(A)}\delta_{\sigma_1(C)\sigma_2(C)}= 4! \cdot 2! = 24\cdot 2, \\
&\sum_{\sigma_1,\sigma_2\in S}\delta_{\sigma_1(A)\sigma_2(A)}\delta_{\sigma_1(C)\sigma_2(C)}\delta_{\sigma_1(G)\sigma_2(G)}= 4! \cdot 1! = 24\cdot 1,\\
&\sum_{\sigma_1,\sigma_2\in S}\delta_{\sigma_1(A)\sigma_2(A)}\delta_{\sigma_1(C)\sigma_2(C)}\delta_{\sigma_1(G)\sigma_2(G)}\delta_{\sigma_1(T)\sigma_2(T)}= 4! \cdot 0! =24,\nonumber
\eqn
which we use to expand the above expressions and carefully collect terms.
This process gives the cubic
\beqn
f^{(13,24)}=\frac{\lambda}{4^5}\left[24 \cdot 24 \cdot 1 (\fra{a}{3})^3+ 24\cdot 6\cdot 3(\fra{a}{3})^2(1-\fra{4a}{3})+24\cdot 2\cdot 3 (\fra{a}{3})(1-\fra{4a}{3})^2+24 \cdot 1 \cdot 1(1-\fra{4a}{3})^3\right]\nonumber,
\eqn
and, noting that $\delta_{ij}^2=\delta_{ij}$, the quintic
\beqn
f^{(14,23)}(P)&=\frac{\lambda}{4^5}\left[24 \cdot 24 \cdot 1(\fra{a}{3})^5+24\cdot 6\cdot 5(\fra{a}{3})^4(1-4\fra{4a}{3})+(24\cdot 6 \cdot 1+24\cdot 2\cdot 9)(\fra{a}{3})^3(1-\fra{4a}{3})^2+\right.\\
&\hspace{3.5em}(24\cdot 2\cdot 3+24\cdot 1\cdot 7)(\fra{a}{3})^2(1-\fra{4a}{3})^3+(24\cdot 1\cdot 3+24\cdot 2)(\fra{a}{3})(1-\fra{4a}{3})^4+\\
&\hspace{29em}\left.24 \cdot 1(1-\fra{4a}{3})^5\right].\nonumber
\eqn
Taking the difference gives
\beqn
q_1(P)=\lambda\frac{1}{2^73^2}a(3^4-225a+276a^2-154a^3+32a^4),\nonumber
\eqn
as required.

\bibliographystyle{sysbio}
\bibliography{squangles}

\end{document}